\documentclass[amsmath,amssymb,aps,prd,a4paper,onecolumn,notitlepage]{revtex4-1}

\usepackage{graphicx}
\usepackage{subfigure}
\usepackage{amsmath}
\usepackage[section]{placeins}
\usepackage{slashed}
\usepackage{color}
\usepackage{soul}
\usepackage{appendix}
\usepackage{placeins}
\usepackage{float}
\usepackage[utf8]{inputenc}
\usepackage{pstricks}
\usepackage{multirow}
\usepackage{longtable}
\usepackage{hyperref}
\usepackage{subfigure}
\usepackage{epsfig}
\usepackage{braket}
\usepackage[normalem]{ulem} 
\usepackage{longtable,booktabs}
\newcommand{\be}{\begin{equation}}
\newcommand{\ee}{\end{equation}}
\newcommand{\ben}{\begin{eqnarray}}
\newcommand{\een}{\end{eqnarray}}

\usepackage{stackengine}

\usepackage{verbatim}
\usepackage{hyperref}
\usepackage{bm}
\usepackage{tikz}
\usetikzlibrary{tikzmark}

\begin{document}


\title{Traces of the $X(3960)$ state in the femtoscopic $D_s^+ D_s^- $ correlations}

\author{Hao-Nan Liu}
\affiliation{School of Physics, Beihang University, Beijing 102206, China}

\author{Zhi-Wei Liu}
\email[Corresponding author: ]{liuzw@buaa.edu.cn}
\affiliation{School of Physics, Beihang University, Beijing 102206, China}

\author{Luciano Abreu}
\email[Corresponding author: ]{luciano.abreu@ufba.br}
\affiliation{Instituto de Física, Universidade Federal da Bahia, Campus Ondina, Salvador, Bahia 40170-115, Brasil}

\author{Li-Sheng Geng}
\email[Corresponding author: ]{lisheng.geng@buaa.edu.cn}
\affiliation{Sino-French Carbon Neutrality Research Center, \'Ecole Centrale de P\'ekin/School of General Engineering, Beihang University, Beijing 100191, China}
\affiliation{School of Physics, Beihang University, Beijing 102206, China}
\affiliation{Peng Huanwu Collaborative Center for Research and Education, Beihang University, Beijing 100191, China}
\affiliation{Beijing Key Laboratory of Advanced Nuclear Materials and Physics, Beihang University, Beijing 102206, China }
\affiliation{Southern Center for Nuclear-Science Theory (SCNT), Institute of Modern Physics, Chinese Academy of Sciences, Huizhou 516000, China}

\begin{abstract}

The femtoscopic $ D_s^+D_s^-$ correlations are investigated to predict the signature of the not-yet-established $X(3960)$ state reported by the LHCb Collaboration, in three scenarios: resonant, virtual, or bound. In the last two scenarios, it might also be identified as the state $X(3930)$. The formalism employed to generate this structure dynamically is based on the Bethe-Salpeter equation with a general $S$-wave potential. We investigate how the relevant properties and observables characterizing this state — such as the pole position, scattering length, and effective range — might be affected by variations in the model parameters. The amplitudes encoding the distinct interpretations of the $X(3960)$ state are then used as input to calculate the femtoscopic correlation function of the $D_s^+ D_s^- $ pair, which is analyzed and discussed.

\date{}
\end{abstract}

\maketitle

\section{\label{sec:intro}Introduction}

\bigskip

Recently, the LHCb Collaboration reported the observation of the so-called $X(3960)$ state in the $D_s^+D_s^-$ invariant mass spectrum of $B^+ \to D_s^+D_s^-K^+$ decays~\cite{LHCb:2022aki}. The analysis favored $J^{PC} = 0^{++}$ quantum numbers, with the following fitted mass and width:
\begin{equation}
m_{\text{X(3960)}} = 3956 \pm 5 \pm 10~\text{MeV}, \quad \Gamma_{\text{X(3960)}} = 43 \pm 13 \pm 8~\text{MeV}.
\label{X3960}
\end{equation}
It appears in addition to another $0^{++}$ state denoted as $X(3930)$, observed by LHCb in the $D^+D^-$ invariant mass distribution of the $B^+ \to D^+D^-K^+$ decay~\cite{LHCb:2020bls,LHCb:2020pxc}, with the following measured observables:
\begin{equation}
m_{\text{X(3930)}} = 3924 \pm 2~\text{MeV}, \quad \Gamma_{\text{X(3930)}} = 17 \pm 5~\text{MeV}.
\label{X3930}
\end{equation}
These LHCb measurements~\cite{LHCb:2020bls,LHCb:2020pxc} are consistent with earlier observations by the Belle~\cite{Belle:2004lle,Belle:2009and} and BaBar~\cite{BaBar:2007vxr,BaBar:2012nxg} Collaborations of a similar state,  referred to as $X(3915)$.

The proximity of both $X(3930)$ and $X(3960)$ to the $D_s^+D_s^-$ threshold ($M_{D_s^+D_s^-}^{\text{thr}} = 3936.7$~MeV) has given rise to intense debate about their possible molecular interpretations, particularly since their masses deviate significantly from quark model predictions for the conventional $\chi_{c0}(3P)$ charmonia~\cite{Barnes:2005pb,Li:2009zu}. Equally important is the controversy concerning whether they represent distinct states or different manifestations of the same structure.
As discussed in Refs.~\cite{Bayar:2022dqa,Ji:2022uie,Ji:2022vdj,Abreu:2023rye}, the near-threshold enhancement in $D_s^+D_s^-$ does not necessarily imply a new resonance. In Ref.~\cite{Bayar:2022dqa}, a coupled-channel analysis of $D\bar{D}$ and $D_s\bar{D}_s$ systems revealed two poles: one below the $D\bar{D}$ threshold, identified as $X(3700)$ (previously predicted in other works~\cite{Gamermann:2006nm,Nieves:2012tt,Hidalgo-Duque:2012rqv,Prelovsek:2020eiw,Brandao:2023vyg,Abreu:2025jqy}), and another below the $D_s^+D_s^-$ threshold corresponding to $X(3930)$, which couples more strongly to $D_s\bar{D}_s$ than to $D\bar{D}$. These couplings successfully reproduce both the $D^+D^-$ peak in $B^+ \to D^+D^-K^+$ decays and the $D_s^+D_s^-$ threshold enhancement in $B^+ \to D_s^+D_s^-K^+$ decays observed by LHCb. In this picture, the $X(3960)$ signal arises from a kinematic threshold enhancement driven by the $X(3930)$ state, interpreted as a molecular state dominated by $D_s\bar{D}_s$ components near threshold. Complementary support comes from Ref.~\cite{Abreu:2023rye}, which employed the same approach to analyze the $J/\psi\omega$ mass distribution in $B^- \to K^- J/\psi\omega$ decays, with the final state arising from rescattering of $D\bar{D}$ and $D_s\bar{D}_s$ intermediate components. 

However, the precise nature of the pole associated with $X(3930)$ remains ambiguous. Ref.~\cite{Ji:2022vdj} utilized an effective field theory based on heavy quark spin symmetry and concluded that the LHCb data can be equally well described by different pole structures: either a bound or virtual state below the $D_s\bar{D}_s$ threshold. The analysis suggested that current experimental data cannot definitively discriminate between these scenarios. Meanwhile, the LHCb amplitude analysis~\cite{LHCb:2022aki} also allows for a conventional resonant state interpretation.

Other approaches have been proposed to describe the $D_s^+ D_s^-$ enhancement. These include molecular interpretations using QCD sum rules~\cite{Xin:2022bzt,Mutuk:2022ckn}, boson exchange models with coupled channels that find resonant states~\cite{Chen:2022dad}, and tetraquark interpretations involving scalar diquark-antidiquark configurations within QCD sum rules~\cite{Agaev:2022pis} or chromomagnetic interaction models~\cite{Guo:2022crh}. The $X(3915)$ and $X(3960)$ have also been proposed as four-quark states within an extended recoupling model~\cite{Badalian:2023qyi}.
Additionally, the production rate of $B^+ \to X(3960) K^+$ has been calculated assuming $X(3960)$ as a molecular $D_s^+ D_s^-$ state using the compositeness condition~\cite{Xie:2022lyw}.

The persistent ambiguity in interpreting the near-threshold $D_s^+ D_s^-$ enhancement highlights the need for additional experimental and theoretical studies employing observables capable of distinguishing the different interpretations. In this regard, femtoscopic correlation analyses offer a promising approach~\cite{Lisa:2005dd,ALICE:2021cpv,Liu:2024uxn,Liu:2025rci}. By measuring two-particle correlations in high-multiplicity collisions, one can extract low-energy interaction parameters such as the scattering length $a_0$ and effective range $r_0$, which are sensitive to the underlying structure of hadronic states~\cite{Chizzali:2022pjd,Liu:2022nec,Liu:2023uly,Liu:2023wfo,Molina:2023oeu,Liu:2024nac,Feijoo:2024bvn,Abreu:2024qqo,Abreu:2025jqy,Liu:2025oar,Shen:2025qpj,Xie:2025xew}. Recent advances in heavy-flavor femtoscopy at LHC suggest that such measurements are becoming feasible~\cite{ALICE:2022enj,ALICE:2024bhk}. A combined analysis of line shapes and femtoscopic correlations could provide complementary constraints, enhancing our understanding of exotic structures~\cite{Liu:2024uxn,Liu:2024nac}.

Thus, in this work, we employ femtoscopic correlations to discriminate between three possible scenarios for the nature of the $X(3960)$ state: a near-threshold resonance, a virtual state, or a bound state. Following the approach of Ref.~\cite{Liu:2024nac}, we utilize the Bethe-Salpeter formalism to parameterize the strong $D_{s}^+ D_{s}^-$ interaction, as detailed in Section~II. The resulting scattering amplitude is then combined with the Coulomb interaction to construct the complete amplitude used in the femtoscopic correlation function (CF), which is presented and analyzed in Section~III. We demonstrate that low-momentum correlation functions across different collision systems (e.g., $pp$, $pA$, and $AA$) exhibit distinctive patterns in each of the three scenarios, thereby providing an alternative tool to resolve the current interpretive ambiguities. Finally, a summary of our main findings and conclusions is presented in Section~IV.


\section{\label{sec:scatt_ampl} The strong interaction contribution to the transition amplitude}

\subsection{Formalism}

We start by introducing the strong contribution to the $D_{s}^+  D_{s}^-$ amplitude, to be obtained via the Bethe-Salpeter formalism and used as input in the $D_{s}^+  D_{s}^-$ CF. 
Here we benefit from the analyses done in Refs.~\cite{Ji:2022uie,Liu:2024nac} exploring distinct configurations for exotic states (see also Refs.~\cite{Feijoo:2023sfe,Dai:2023cyo}). In particular, Ref.~\cite{Ji:2022uie} employed a contact potential from a nonrelativistic effective field theory considering heavy quark spin and light flavor SU(3) symmetries. The low-energy constants (LECs) have been determined by fitting the line shapes of the $D_{s}^+ D_{s}^-$ invariant mass distribution of the $B^+ \to D_{s}^+ D_{s}^- K^+$ reaction to the LHCb data reported in Ref.~\citep{LHCb:2022aki}. The conclusion of Ref.~\cite{Ji:2022uie} was that the $X(3960)$ state might be compatible with either a bound or virtual state in the $D_{s}^+  D_{s}^-$ channel. 
By its turn, in the study of the possible femtoscopic signatures of the $Z_c (3900)/ Z_{cs}(3985)$ states, Ref.~\citep{Liu:2024nac} used a general $S$-wave potential up to the next-to-leading order in momentum expansion for the $D^0 D^{\ast -}/D^0 D_s^{\ast -}$ systems. Accordingly, the LECs have been determined by exploring the scenarios of bound, virtual, and resonant states. In the latter case, the poles reported by the PDG have been employed, and the former two have been fine-tuned to obtain line shapes similar to that of the resonant state. 
For the determination of the strong $D_{s}^+  D_{s}^-$ amplitude, we adopt the potential and fit the LECs similarly to the framework of~\citep{Liu:2024nac}.

We define the $S$-wave projected strong potential for the process $D_{s}^+  D_{s}^- \to D_{s}^+  D_{s}^-$ as 
\begin{align}
 V^\textrm{(S)} = a + b \ k^2 ,\label{Vpot}
\end{align}
where  $k = \lambda^{1/2} (s,m_{D_s}^2,m_{D_s}^2)/(2\sqrt{s})$ is the relative momentum of the meson pair, with $\lambda (x, y, z) $ being the K\"allen function, $m_{D_s}$ the mass of the $D_{s}^+,  D_{s}^-$  mesons and $s$ the squared CM energy; and $a, b$ are the free parameters (i.e., LECs) to be determined. 

The unitarized scattering amplitude matrix of the strong interaction $T^\textrm{(S)}$ is then obtained by solving the Bethe-Salpeter equation, 
\begin{align}
    T^\textrm{(S)} = \left[1 -V^\textrm{(S)}G\right]^{-1}V^\textrm{(S)},
\label{BSE}
\end{align}
where $G$ is the loop function for two intermediate mesons in a given channel, which in the cutoff regularization scheme can be written as
\begin{equation}
G \left(\sqrt{s}\right)= \int_0^{q_{max}} \frac{q^2 {\rm d}q}{2\pi^2} \frac{1}{4 \omega^2_{D_s} (q) } \frac{1}{\sqrt{s} - 2\omega_{D_s}(q) + i \varepsilon }~,
\label{gfn}
\end{equation}
with $\omega_{D_s} (q) = \sqrt{q^2 + m_{D_s}^2}$, and  $q_{max}$ being the cutoff chosen within the range $0.5 - 1.0 $ GeV. 

We identify virtual, bound, and resonant states through poles of the $T^\textrm{(S)}$-matrix in Riemann sheets of the $\sqrt{s}$ plane~\cite{Liu:2024nac}. Virtual and bound states correspond to poles on the real axis below threshold in the second and first Riemann sheets, respectively;  while resonances appear as complex poles in the second Riemann sheet, with their mass and half-width being given by the real and imaginary parts of the pole position. The second Riemann sheet loop function $G^{(II)}$ is obtained via analytic continuation: $G^{(II)} = G + 2i\rho$, where $\rho = k/(8\pi\sqrt{s})$ is the phase space factor.

We also calculate two key low-energy scattering observables defined by the effective range expansion~\cite{Liu:2024nac,Dai:2023cyo,Feijoo:2023sfe,Khemchandani:2023xup,Shen:2024npc}: the scattering length $a_0$ and the effective range $r_0$. Using the relation between the $T^\textrm{(S)}$-matrix and phase shift $\delta_0$, $T^\textrm{(S)} \propto (k \cot \delta_0 - i k)^{-1}$, the expansion can be written as 
\begin{align}
-8\pi \sqrt{s} \ T^{\textrm{(S)}-1}  + i k  = -\frac{1}{a_0} + \frac{1}{2} r_0 k^2 + \mathcal{O}(k^4). \label{eff_range_exp}
\end{align}
From this relation, $a_0$ and $r_0$ are given by 
\begin{align}
a_0 &= \left. \frac{T^\textrm{(S)}}{8\pi \sqrt{s}} \right|_{s=s_{\mathrm{thr}}}, \nonumber \\ 
 r_{0} &=\left. \frac{\partial^2 }{\partial k^2 } \left(-8\pi \sqrt{s} \ {T^\textrm{(S)}}^{-1} + ik \right) \right|_{s=s_{\textrm{thr}}}. \label{alength}
\end{align}
Here, $s_{\mathrm{thr}}$ is the squared CM energy at the channel's threshold.

To determine the LECs, we employ the $D_{s}^+ D_{s}^-$ invariant mass distribution of the $B^+ \to D_{s}^+ D_{s}^- K^+$ reaction, which is given by~\cite{Ji:2022uie,Bayar:2022dqa} 
\begin{align}
  \frac{d\Gamma}{dM_{\rm inv}} = \frac{1}{(2\pi)^3}\frac{1}{4 m^2_{B}} p_{K^+} \tilde{p}_{D_s} | \widetilde{T} |^2,
      \label{distr_mass}
\end{align}
where $M_{\rm inv}$ is the invariant mass of the  $D_{s}^+ D_{s}^-$ system; $p_{K^+} = \lambda^{1/2}(m_{B^+}^2, m^2_{K^+}, M_{\rm inv}^2)/2m_{B^+}$ and  $\tilde{p}_{D_s} = \lambda^{1/2}(M_{\rm inv}^2, m^2_{D_{s}^+},$ $ m^2_{D_{s}^-})/2M_{\rm inv}$ are respectively the $K^+$ and $D_{s}$ momenta; and $\tilde{T}$ is the amplitude for the $B^+ \to D_{s}^+ D_{s}^- K^+$ transition.

\begin{figure}[htbp!]  
  \centering
  \includegraphics[width=6.5cm]{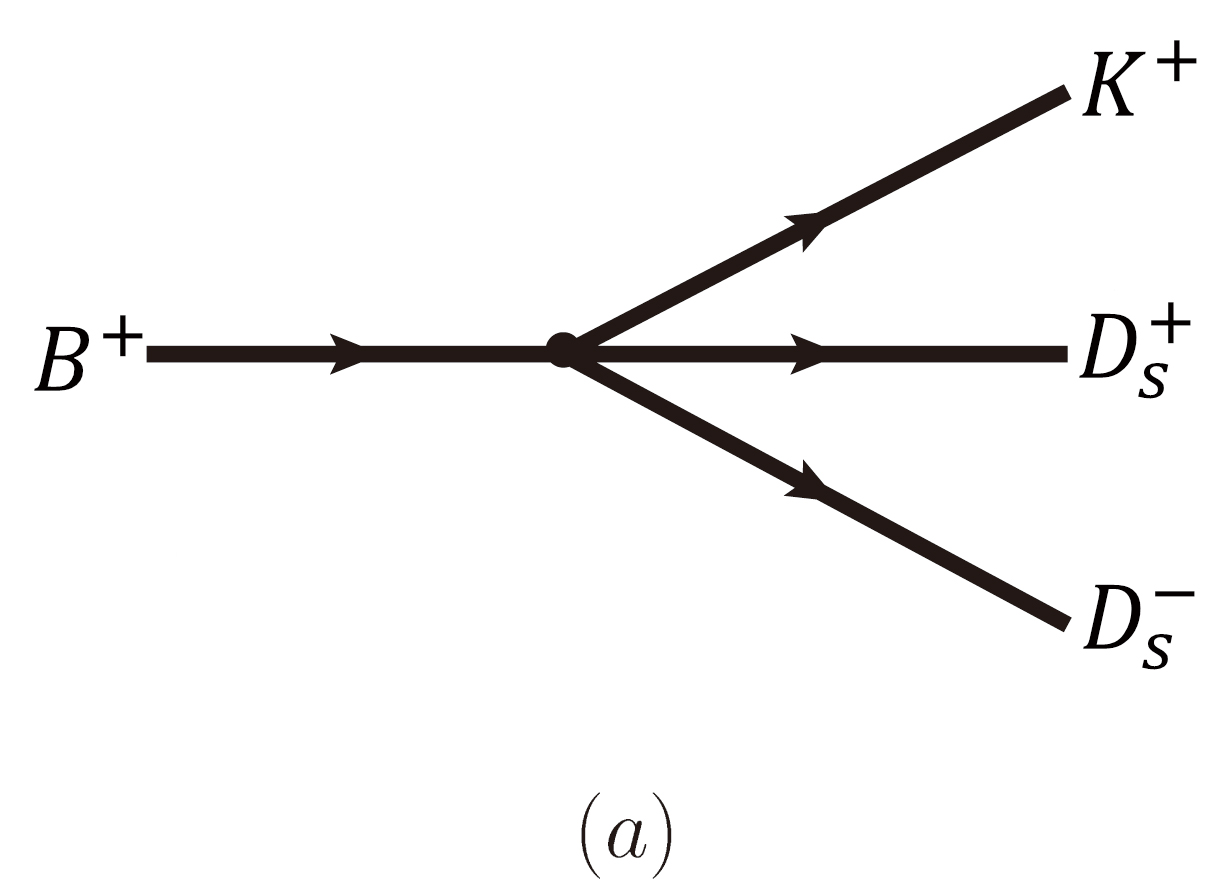}
  \includegraphics[width=7.5cm]{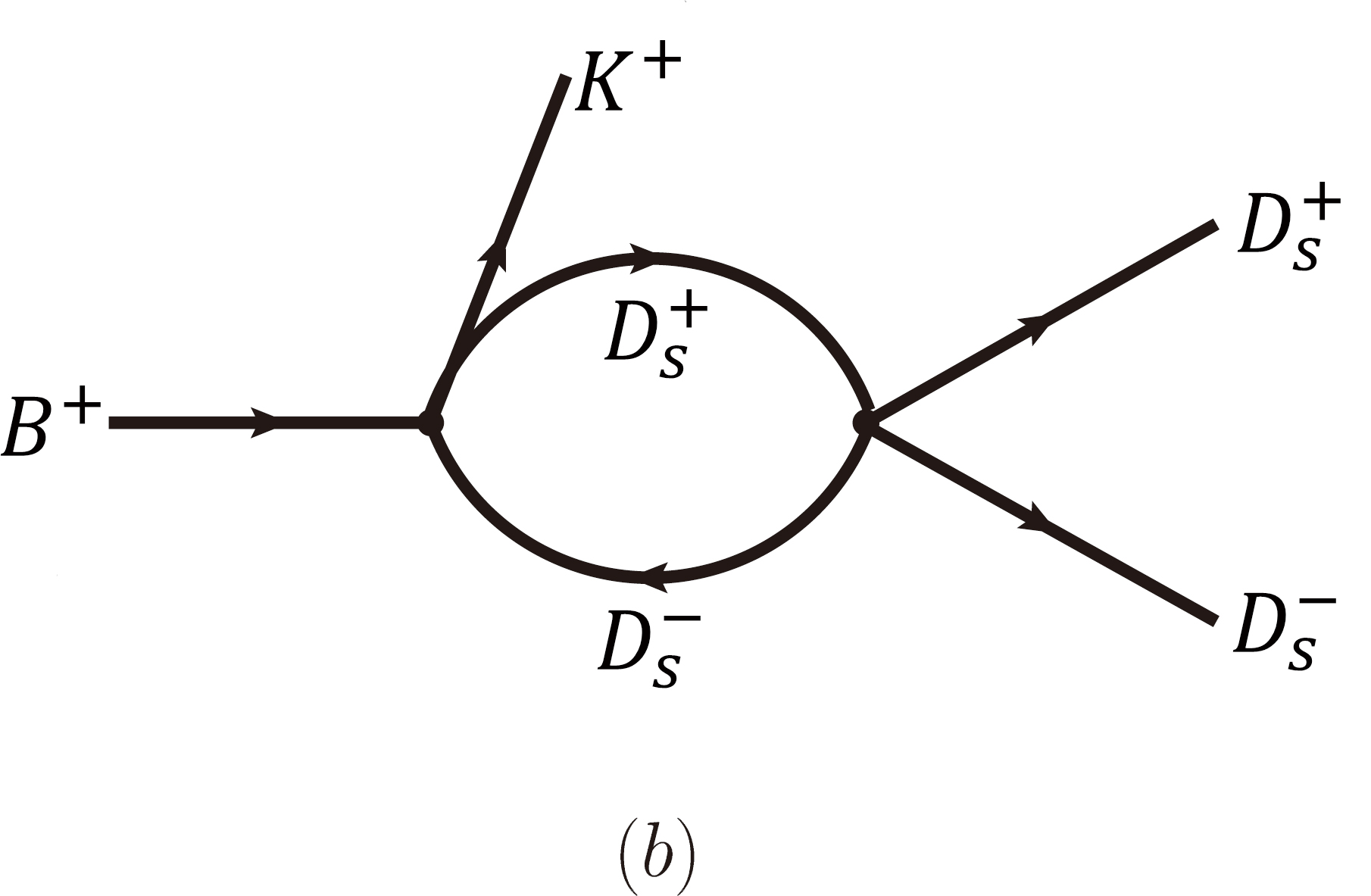}
   \caption{Mechanisms contributing to the $B^+ \to D_{s}^+ D_{s}^- K^+$ reaction. (a) Tree-level contribution. (b) Rescattering contribution. }
  \label{fig1}
\end{figure}

The analytical expression for $\widetilde{T}$ can be obtained as follows~\cite{Ji:2022uie,Bayar:2022dqa}. At the quark level, these $B$-decays proceed via a Cabibbo-suppressed internal $W$-emission: $\bar b (\to \bar c W^+ \to \bar c c \bar s) u \to \bar c c (\bar s u)$. After hadronization of the $c \bar c$ pair, only the $D_{s}^+ D_{s}^-$ channel remains. At the hadron level, the production mechanisms considered for $B^+ \to D_{s}^+ D_{s}^- K^+$ are the tree-level and one-loop contributions, as depicted in Fig.~\ref{fig1}. Consequently, the effects of the state under investigation are produced through the interaction of the intermediate $D_{s}^+ D_{s}^-$ pair in the rescattering contribution.

So, the amplitude for the diagrams in Fig.~\ref{fig1} is given by
\begin{align}
 \widetilde{T}(M_{\rm inv}) = C \left[ 1 + G(M_{\rm inv}) T^\textrm{(S)} (M_{\rm inv}) \right], \label{tranmatrix}
\end{align}
where $C$ is an overall constant encoding effectively the information of the weak vertex, $G(M_{\rm inv})$ is the two-meson loop function given in Eq.~(\ref{gfn}), and $T^\mathrm{(S)}(M_{\rm inv})$ is the unitarized scattering amplitude matrix given in Eq.~(\ref{BSE}). 

\subsection{Fitting Results}

Using Eq.~(\ref{tranmatrix}) and Eq.~(\ref{distr_mass}), we fit the LECs $a$ and $b$ of the potential $V^\mathrm{(S)}$ to reproduce the LHCb data~\citep{LHCb:2022aki} via the least-squares minimization. In the bound and virtual state scenarios, the fit considers the first ten data points in the CM energy region from 3.92 to 4.12~GeV (data before the third vertical gray line), as in Ref.~\cite{Ji:2022uie}. In the resonant-state scenario, only the first four data points (those before the second vertical gray line) are used. These restricted ranges are chosen because our model focuses on the region near the $D_{s}^+ D_{s}^-$ threshold and excludes other irrelevant effects or states at higher energies. In the resonant state scenario, the invariant mass spectrum is expected to feature a narrow, Breit-Wigner-like peak (as shown in Fig.~2 of Ref.~\cite{LHCb:2022aki}), whose essential characteristics are encapsulated within the first four data points. Fitting exclusively to these points thus results in a small $\chi^2/{\rm d.o.f.}$ value, which we consider sufficient for a qualitative discussion of the resonant interpretation.

\begin{figure}[htbp!]  
  \centering
    \subfigure[]{
  \includegraphics[width=8cm]{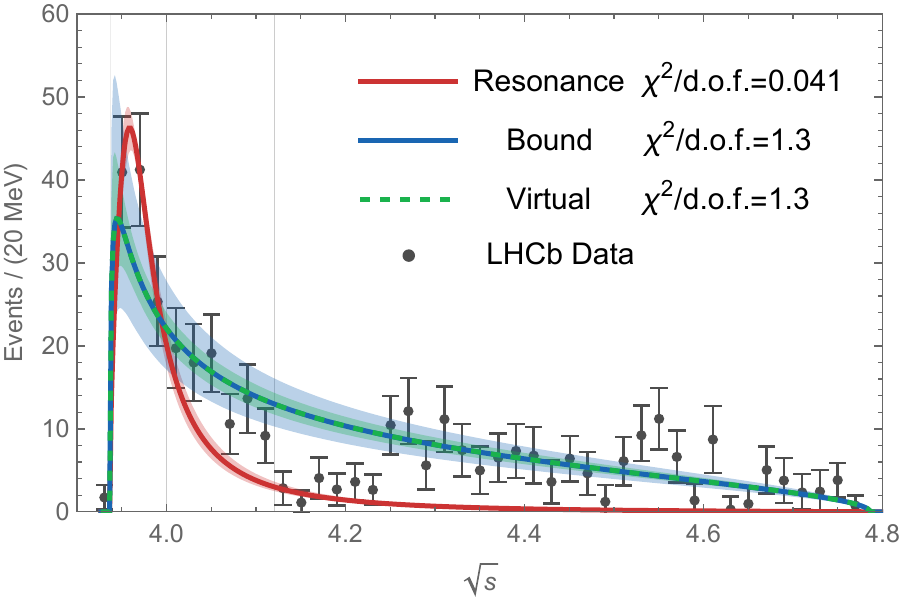}
\label{fig_mass_distr-a}
}
    \subfigure[]{
\includegraphics[width=8cm]{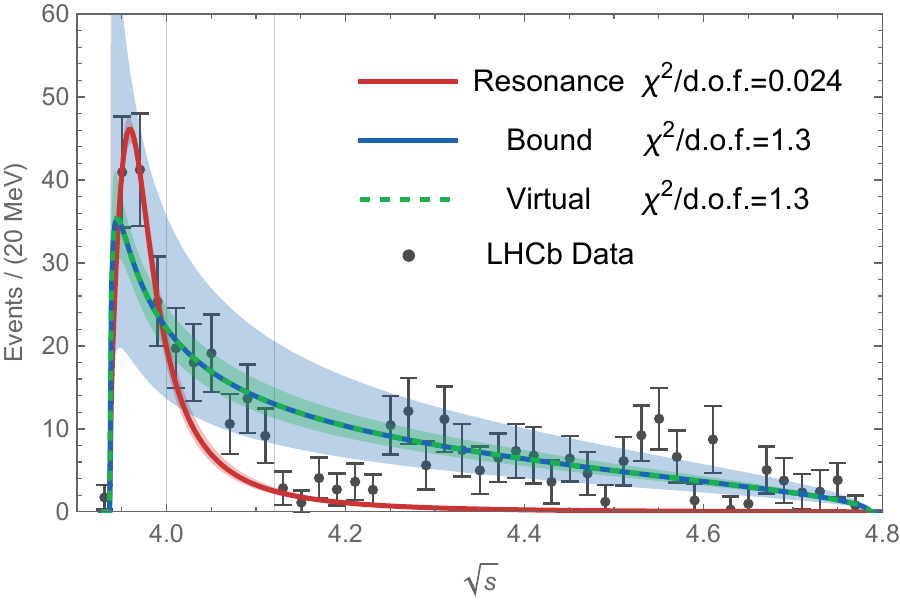}
\label{fig_mass_distr-b}
}
\caption{Differential distributions of the  $B^+ \to D_{s}^+ D_{s}^- K^+$ decay taking \ref{fig_mass_distr-a} $ q_{\text{max}} = 1.0$~GeV and \ref{fig_mass_distr-b}$  q_{\text{max}} = 0.5$~GeV. The shaded areas represent the uncertainties in the LECs displayed in Table~\ref{table_scatprob} for the different scenarios. The experimental data are taken
from Ref.~\cite{LHCb:2022aki}. The first vertical gray line corresponds to the $D_{s}^+ D_{s}^-$ threshold.}
  \label{fig_mass_distr}
\end{figure}

Fig.~\ref{fig_mass_distr} shows the fitted differential mass distributions for a cutoff parameter $q_{\text{max}} = 1.0$~GeV in the bound, virtual, and resonant scenarios. All three cases produce line shapes compatible with the data, with those for bound and virtual states being practically indistinguishable. Table~\ref{table_scatprob} provides the fitted values of $a$ and $b$, along with the resulting pole positions, scattering lengths $a_0$, and effective ranges $r_0$ in the three scenarios. For completeness, we also include results for $q_{\text{max}} = 0.5$~GeV.

The pole positions for the virtual and bound states interpretations are consistent with those obtained in Refs.~\cite{Ji:2022uie,LHCb:2020pxc} within uncertainties, which are associated with the $X(3930)$ state. The resonant-state scenario yields a pole position consistent with that reported by the LHCb Collaboration~\cite{LHCb:2022aki}, where a new state, referred to as $X(3960)$, is claimed.

\begin{table}[htbp!]
\caption{Relevant quantities obtained from the fits to the
LHCb data in Ref.~\citep{LHCb:2022aki}: mass and width of the pole position, the LECs $a, b$, the scattering length $a_0$, and the effective range $r_{0}$. The $D_s^+ D_s^-$ threshold is 3936.7 MeV.  }\label{table_scatprob}
\begin{ruledtabular}
\begin{tabular}{ccccccc}
\multicolumn{7}{c}{$q_{max}= 1.0 $ GeV} \\
\hline
Scenario & $ M $  [MeV] & $ \Gamma $  [MeV]  & $a$ & $b \ [10^{-6} \ \mathrm{MeV}^{-2}]$  & $a_{0}$  [fm] & $r_{0}$  [fm] \\
\hline
Resonance & $ {3948.79}^{+2.44}_{-3.33} $ & $ {60.04}^{+10.83}_{-8.72} $ & $-106.67 \pm 3.53$ & $-629.75 \pm 73.62$   & $ {-0.63}^{+0.06}_{-0.07} $  & $ {-1.86}^{+0.28}_{-0.30} $    \\
Bound & $ {3928.27}^{+3.55}_{-4.07} $ & 0 & $-202.61 \pm 11.22$ & 0.00  & $ {1.60}^{+0.48}_{-0.28} $  & $ {0.28}^{+0.001}_{-0.001} $    \\
Virtual & $ {3928.28}^{+3.31}_{-4.44} $ & 0 & $-132.57 \pm 5.67$ & 0.00  & $ {-1.47}^{+0.29}_{-0.43} $  & $ {0.29}^{+0.002}_{-0.002} $    \\
\hline
\hline
\multicolumn{7}{c}{$q_{max}= 0.5 $ GeV} \\
\hline
Scenario & $ M $  [MeV] & $ \Gamma $  [MeV]  & $a$ & $b \ [10^{-6} \ \mathrm{MeV}^{-2}]$  & $a_{0}$  [fm] & $r_{0}$  [fm] \\
\hline
Resonance & $ {3949.43}^{+2.04}_{-2.60} $ & ${63.75}^{+8.30}_{-6.98}$  & $-136.79 \pm 8.17$ & $-1431.33 \pm 145.39$  & ${-0.48}^{+0.05}_{-0.05}$  & ${-2.45}^{+0.44}_{-0.51}$    \\
Bound & ${3928.27}^{+3.85}_{-3.73} $   & 0 & $-517.19 \pm 73.08$ & 0.00  & ${1.60}^{+0.53}_{-0.26} $  & ${0.51}^{+0.002}_{-0.001}$    \\
Virtual & ${3928.28}^{+3.24}_{-4.57} $ & 0 & $-220.19 \pm 15.64$ & 0.00 & ${-1.47}^{+0.30}_{-0.42} $  & ${0.53}^{+0.002}_{-0.002}$    \\
\end{tabular}
\end{ruledtabular}
\end{table}

\begin{figure}[htbp!]
    \centering
    \subfigure[]{
    \includegraphics[width=0.45\textwidth]{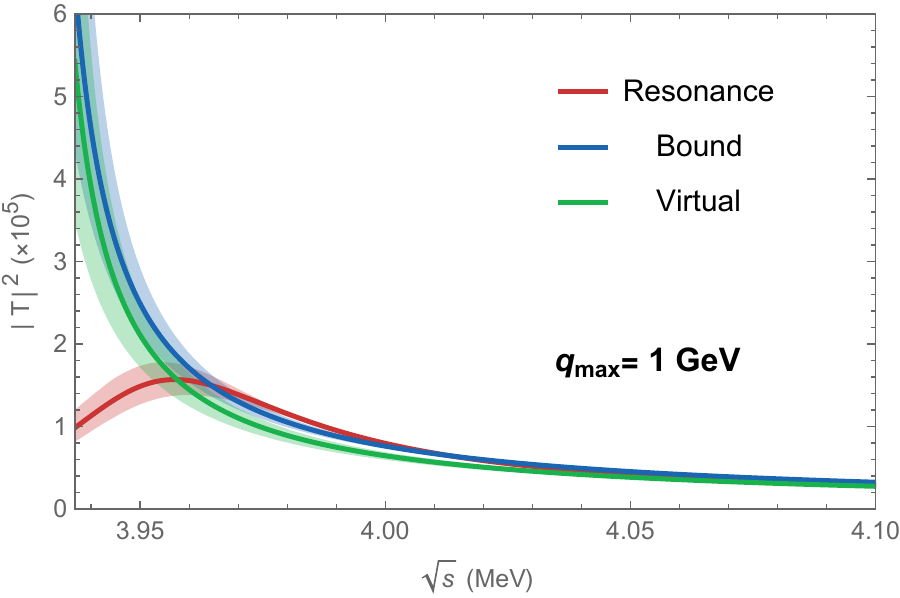}
    \label{fig_squared_ampl1}
    }
    \subfigure[]{
    \includegraphics[width=0.45\textwidth]{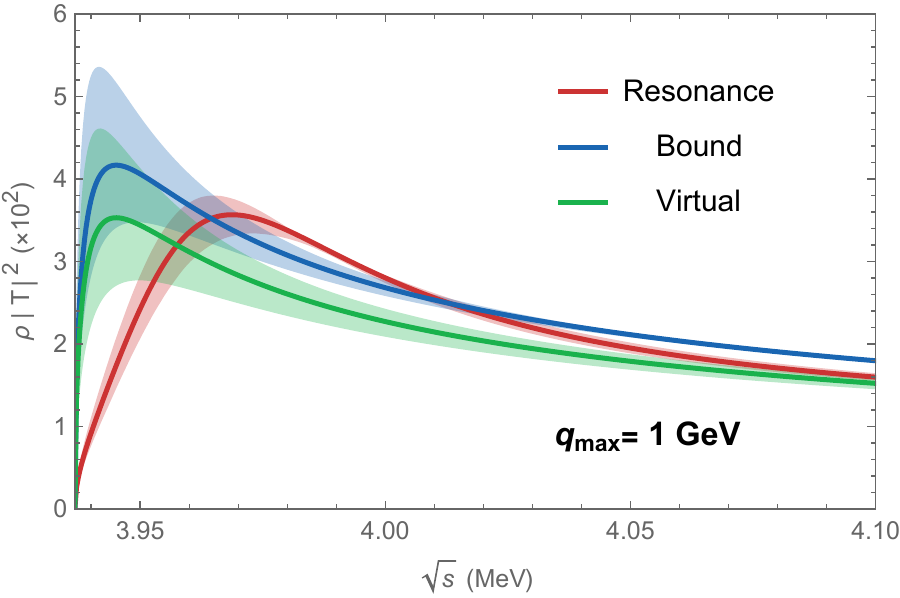}
    \label{fig_rho_squared_ampl1}
    }
\caption{\ref{fig_squared_ampl1}: modulus square of the amplitude of the $ D_s^+ D_s^- \rightarrow  D_s^+ D_s^- $ channel as a function of the CM energy. \ref{fig_rho_squared_ampl1}: modulus square of the amplitude times the phase-space factor of the $ D_s^+ D_s^- \rightarrow  D_s^+ D_s^- $ channel as a function of the CM energy. The shaded areas represent the uncertainties in the LECs displayed in Table~\ref{table_scatprob} for the different scenarios.  
}
\label{fig_ampl}
\end{figure}

For completeness, Fig.~\ref{fig_ampl} shows the squared amplitude and the squared amplitude times the phase-space factor for the $D_s^+ D_s^- \rightarrow D_s^+ D_s^-$ channel as a function of the CM energy, for a cutoff parameter $q_{\text{max}} = 1.0$~GeV. 
The inherent kinematic suppression due to the phase-space factor in the near-threshold region makes it difficult to distinguish among the different configurations in the invariant-mass spectrum. 

It is most important to emphasize that all three scenarios can fairly reproduce the low-energy experimental data within current uncertainties. This demonstrates that the invariant mass distribution alone is insufficient to distinguish between these interpretations. Additional observables are therefore necessary to discriminate among the different configurations and clarify this ambiguity.

We now examine how these scenarios manifest in femtoscopic correlation functions.

\section{\label{sec-cf}Correlation Functions}

\subsection{Formalism}

The correlation function (CF) for two-particle systems is defined as the ratio of the probability of measuring the two-particle state and the product of the probabilities of measuring each particle~\cite{Fabbietti:2020bfg}. According to the Koonin–Pratt approach, after some manipulations, the CF can be given by the following formula ~\cite{Lisa:2005dd,Koonin:1977fh,Pratt:1986cc}
\begin{eqnarray}
C(k) & = & \int d^3 r \ S_{12}(\bm{r}) \ \vert  \Psi (\bm{k} ; \bm{r}) \vert ^2, 
\label{cf1}
\end{eqnarray}
where $ \bm{k} $ is the relative momentum in the CM frame of the pair; $\bm{r}$ is the relative distance between the two particles; $\Psi (\bm{k} ; \bm{r})$ is the relative two-particle wave function carrying information of final-state interactions; and $ S_{12}(\bm{r}) $ is the source function encoding the particle-emitting source. 

In the present work, we adopt the source function in its usual static Gaussian profile normalized to unity, i.e., ~\cite{Lisa:2005dd,ALICE:2021cpv}
\begin{eqnarray}
S_{12}(\bm{r})  & = & \frac{1}{\left(  4 \pi \right)^{\frac{3}{2}} R^3} \exp{\left(  -\frac{r^2}{4 R^2 }\right)}, 
\label{sourcef1}
\end{eqnarray}
where $R$ is the source size parameter, chosen in such a way to reproduce high-multiplicity events in pp collisions (smaller values) to central AA collisions (larger values).

A realistic treatment of the $\left\vert D_s^+ D_s^- \right\rangle$ system must include the Coulomb interaction due to its significant role~\cite{Ge:2025put}. We therefore employ a relative wave function that incorporates both the $S$-wave strong interaction and the Coulomb potential~\cite{Encarnacion:2024jge}:
\begin{align}
   \Psi (\bm{k} ; \bm{r}) = \Phi^{\mathrm{(C)}}(\bm{k} ; \bm{r}) - \Phi^{\mathrm{(C)}}_0(k,r) + \psi_0(k,r),
   \label{reL_wf1}
\end{align}
where $\Phi^{\mathrm{(C)}}(\bm{k} ; \bm{r})$ is the complete Coulomb wave function~\cite{joachain1975quantum}, $\Phi^{\mathrm{(C)}}_0(k,r)$ is its $S$-wave component, and $\psi_0(k,r)$ contains the full strong interaction in the presence of the Coulomb potential.

Using the wave function in Eq.~(\ref{reL_wf1}) and assuming a spherically symmetric source function from Eq.~(\ref{sourcef1}), the Koonin–Pratt formula in Eq.~(\ref{cf1}) becomes~\cite{Vidana:2023olz,Feijoo:2023sfe,Albaladejo:2023pzq,Khemchandani:2023xup,Abreu:2024qqo,Ge:2025put,Liu:2022nec}
\begin{eqnarray}
C(k) & = & \int \mathrm{d}^3\bm{r} \, S_{12}(\bm{r}) \, \left\vert \Phi^{\mathrm{(C)}}(\bm{k} ; \bm{r}) \right\vert^2 + \int_{0}^{\infty} 4 \pi r^2 \mathrm{d} r \, S_{12}(\bm{r}) \left( \left\vert \psi_0(k,r) \right\vert^2 - \left\vert \Phi^{\mathrm{(C)}}_0(k,r) \right\vert^2 \right). 
\label{cf3}
\end{eqnarray}

Following the notation of Refs.~\cite{Torres-Rincon:2023qll,Abreu:2025jqy}, the complete Coulomb wave function is given by~\cite{joachain1975quantum}
\begin{equation}
    \Phi^{\mathrm{(C)}}(k; \bm{r},z)= e^{-\pi \gamma/2} \Gamma(1+ i\gamma) e^{ikz} \, {_1}F_{1}(-i\gamma;1;ik(r-z)) . \label{eq:completeCoulomb}
\end{equation}
 Here, $\Gamma(z)$ is the Euler gamma function, ${_1}F_{1}(a;b;z)$ is the confluent hypergeometric function (Kummer's function), and $\gamma$ is the Sommerfeld parameter,
\begin{equation}
\gamma = Z_1 Z_2 \frac{\mu \alpha}{k}, \label{eq:sommerfeld}
\end{equation}
with $Z_1Z_2$ being the product of the charges, $\alpha$ the fine-structure constant, and $\mu$ the reduced mass of the hadron pair.

The $S$-wave function $\psi_0(k,r)$ is obtained from the Lippmann-Schwinger equation $ \vert \psi \rangle = \vert \phi \rangle + G T \vert \phi \rangle$, where $ \vert \phi \rangle$ is the free wave function~\cite{Torres-Rincon:2023qll,Encarnacion:2024jge,Liu:2023uly}:
\begin{align}
    \psi_0(k,r) = j_0(kr) + \int_0^{q_{\mathrm{max}}} \frac{q^2 \mathrm{d}q}{2\pi^2} \frac{1}{4 \omega^2_{D_s}(q)} \frac{T(k,q;s)\, j_0(qr)}{\sqrt{s} - 2\omega_{D_s}(q) + i \varepsilon}. \label{eq:psi0}
\end{align}
Here, $j_{\nu}(x)$ is the spherical Bessel function of order $\nu$, and $T(k,q;s)$ is the $S$-wave scattering amplitude for the charged meson pair including both strong and Coulomb contributions.

Ideally, the amplitude $T(k,q;s)$ should be obtained by simultaneously unitarizing the combined strong and Coulomb amplitudes. Since the Coulomb contribution depends on the meson momenta in both initial and final states, this requires solving the full off-shell Bethe-Salpeter equation. However, following the analysis in Ref.~\cite{Encarnacion:2024jge}, the momentum integral in Eq.~(\ref{eq:psi0}) is calculated with the Coulomb amplitude up
to first order in the Bethe-Salpeter equation combined with the on-shell strong amplitude studied in the previous Section. Given the exploratory nature of this study, this approximation has the practical advantage of leaving the fitted strong amplitude unchanged.

Explicitly, the full scattering amplitude is decomposed as
\begin{align}
    T(p,p';s) = T^\textrm{(S)}(\sqrt{s}) + T^\textrm{(C)}(p,p';s), 
    \label{full_ampl1}
\end{align}
where $T^\textrm{(S)}(\sqrt{s})$ is the unitarized strong amplitude from Eq.~(\ref{BSE}) and $T^\textrm{(C)}(p,p';s)$ is the Coulomb amplitude in the Born approximation:
\begin{align}
     T^\textrm{(C)}(p,p';s) = V^{\textrm{(C,rel)}}_0(p,p';s).
    \label{Coul_ampl}
\end{align}

To obtain the Coulomb contribution $V^{\mathrm{(C,rel)}}_0(p,p';s)$, we begin with the Fourier transform of the potential $V^{(C)} = \epsilon \alpha/r$, where $\epsilon = +1\,(-1)$ for identical (opposite) charged particles, into momentum space:
\begin{align}
    V^\mathrm{(C)}_{\mathrm{total}}(|\bm{p}'-\bm{p}|) = \int_0^{R_C} \mathrm{d}^3 r \, e^{i (\bm{p}'-\bm{p}) \cdot \bm{r}} \frac{\epsilon \alpha}{r} 
    = \frac{4 \pi \epsilon \alpha}{|\bm{p}'-\bm{p}|^2} \left(1 - \cos(|\bm{p}'-\bm{p}| R_C) \right),
    \label{VC-tot}
\end{align}
where only the short-range Coulomb interaction ($r < R_C$) is considered to avoid divergences in the integrals. The long-range behavior will be accounted for through the asymptotic wave functions.

The $S$-wave component of this potential is obtained by projecting it onto the partial wave:
\begin{align}
    V^\mathrm{(C)}_0(p,p') & = \frac{1}{2} \int_{-1}^1 \mathrm{d}(\cos\theta_{\bm{pp}'}) \, V^\mathrm{(C)}_{\mathrm{total}}(|\bm{p}'-\bm{p}|) \\
    & = \frac{2 \pi \epsilon \alpha}{p p'} \left\{ \mathrm{Ci}[|p'-p|R_C] - \mathrm{Ci}[|p'+p|R_C] + \ln\left( \frac{p'+p}{|p'-p|} \right) \right\},
\end{align}
where $\mathrm{Ci}[x] = \int_x^\infty \mathrm{d}t \, (\cos t)/t$ is the cosine integral function. We have tested various values of $R_C$ and find that our results stabilize at $R_C = 4$ fm, which we therefore adopt in all subsequent calculations.

To consistently incorporate the Coulomb interaction within the relativistic Bethe-Salpeter equation, the $S$-wave Coulomb potential must be modified with relativistic kinematic factors~\cite{Torres-Rincon:2023qll,Encarnacion:2024jge}:
\begin{align}
    V^{\mathrm{(C,rel)}}_0(p,p';s) = 2 \omega_{D_s}(p) \sqrt{\xi(p;s)} \, V^\mathrm{(C)}_0(p,p') \, \sqrt{\xi(p';s)} \, 2 \omega_{D_s}(p'),
\end{align}
where the kinematic factors $\xi(p;s)$ are given by:
\begin{align}
    \xi(p;s) = 2\mu \frac{\sqrt{s} - 2 \omega_{D_s}(p)}{\frac{\lambda(s,m_1^2,m_2^2)}{4s} - p^2},
\end{align}
with $\mu = m_{D_s}/2$ being the reduced mass.


\subsection{Results and Discussions}

\begin{figure}[htbp!]
    \centering
    \centering
    \subfigure[]{
    \includegraphics[width=0.45\textwidth]{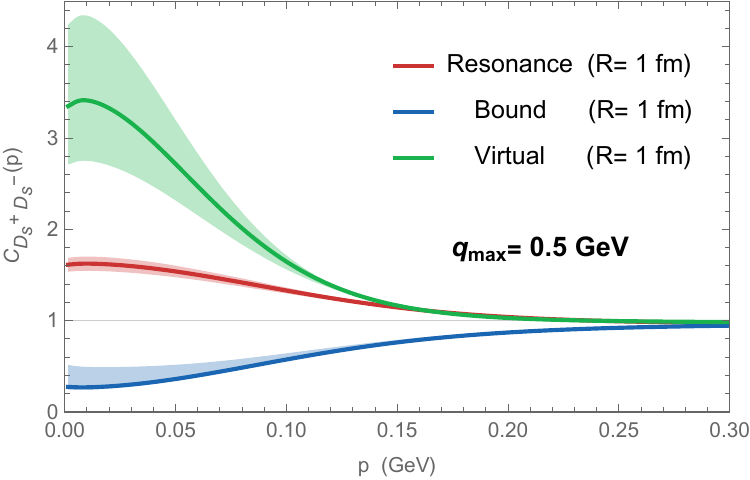}
    \label{fig_CF_strong_1fm_0.5GeV}
    }
    \subfigure[]{
    \includegraphics[width=0.45\textwidth]{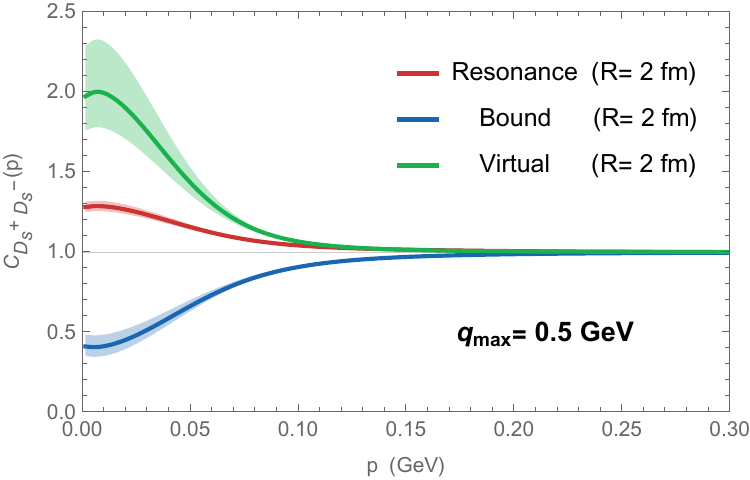}
    \label{fig_CF_strong_2fm_0.5GeV}
    }
     \subfigure[]{
    \includegraphics[width=0.45\textwidth]{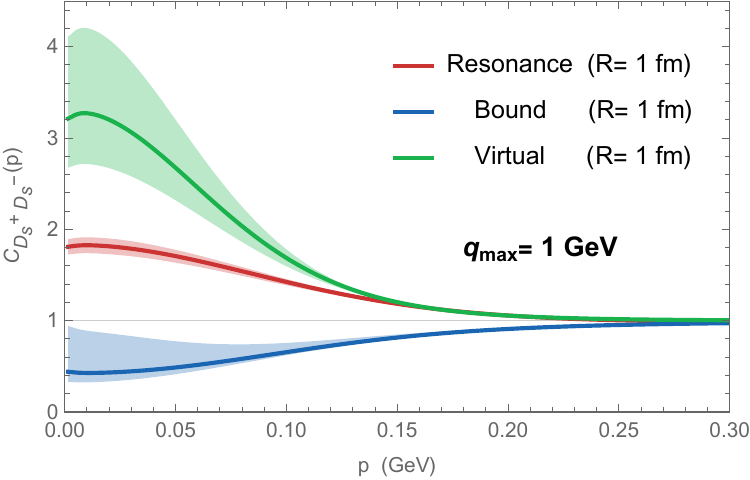}
    \label{fig_CF_strong_1fm_1GeV}
    }
    \subfigure[]{
    \includegraphics[width=0.45\textwidth]{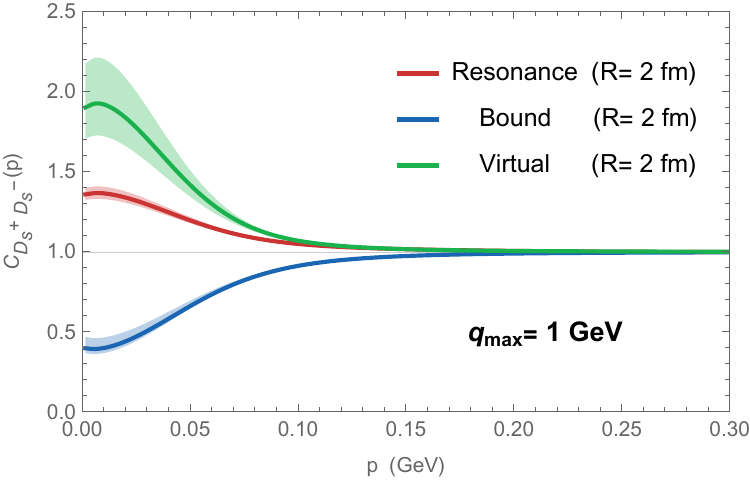}
    \label{fig_CF_strong_2fm_1GeV}
    }
    \caption{\ref{fig_CF_strong_1fm_0.5GeV} - \ref{fig_CF_strong_2fm_1GeV}: the pure strong contribution of the $  D_s^+ D_s^- $ CF as a function of the CM relative momentum $k$, taking different values of the size parameter $R$ in three scenarios. The shaded areas represent the uncertainties in the LECs displayed in Table~\ref{table_scatprob} for the different scenarios.}
    \label{fig_CF_strong}
\end{figure}

Fig.~\ref{fig_CF_strong} shows the pure strong $D_s^+ D_s^-$ CF as a function of the CM relative momentum $k$ for different values of the parameter $R$ in three scenarios. This contribution is calculated by replacing the $\Phi^{\rm (C)}$ and $\Phi^{\rm (C)}_0$ with $e^{i k\cdot r}$ and $j_0$ in Eq.~(\ref{cf3}) and using the scattering amplitude $T(k,q;s) = T^\mathrm{(S)}(s)$ in Eq.~(\ref{eq:psi0}), which contains only the unitarized strong interaction. 
In the low-momentum region, the pure strong $D_s^+ D_s^-$ CF exhibits distinct and typical behavior for each configuration. Specifically, compared to unity, the virtual and bound-state scenarios produce, respectively, enhancements and suppression in the low-momentum correlations, while the resonant configuration yields a moderate enhancement. This distinction is most pronounced for smaller values of the source size parameter $R$ and becomes less significant as $R$ increases. 
These findings are qualitatively consistent with those obtained in Ref.~\cite{Liu:2024nac} in investigating femtoscopic signatures for the $Z_c(3900)/Z_{cs}(3985)$ states in the $D^0 D^{\ast -}/D^0 D_s^{\ast -}$ systems. However, a difference can be noted in the resonant interpretation. Due to the larger width given in Table~\ref{table_scatprob}, the $D_s^+ D_s^-$ CF in Fig.~\ref{fig_CF_strong} does not exhibit a pronounced dip in the intermediate-momentum region. Specifically, in the momentum region slightly above the resonance pole, i.e., $k_R \gtrsim 160$ MeV, the CF has an almost plateaulike appearance,  unlike the cases of the $D^0 D^{\ast -}/D^0 D_s^{\ast -}$ systems studied in Ref.~\cite{Liu:2024nac}.

It is worth noting that although the CF of a bound state exhibits a ``suppression" in the low-momentum region—similar to the behavior observed in systems with repulsive interactions—the underlying physical mechanisms differ significantly. In the bound state scenario, pairs that form a bound state are lost to the correlation yield, leading to a suppression in the low-momentum region of the CF. In contrast, for the repulsive interaction case, the repulsive force accelerates the particles apart. This increases their relative momentum and similarly suppresses it at low momentum in the CF.

\begin{figure}[htbp!]
    \centering
    \centering
    \subfigure[]{
    \includegraphics[width=0.45\textwidth]{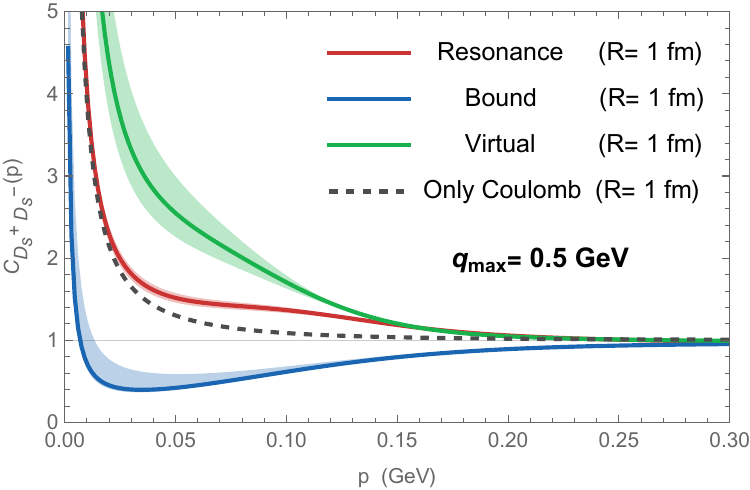}
    \label{fig_CF_total_1fm_0.5GeV}
    }
    \subfigure[]{
    \includegraphics[width=0.45\textwidth]{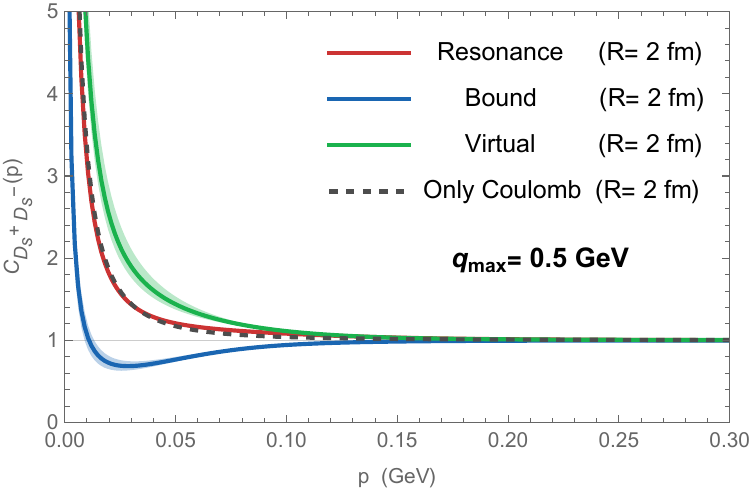}
    \label{fig_CF_total_2fm_0.5GeV}
    }
    \subfigure[]{
    \includegraphics[width=0.45\textwidth]{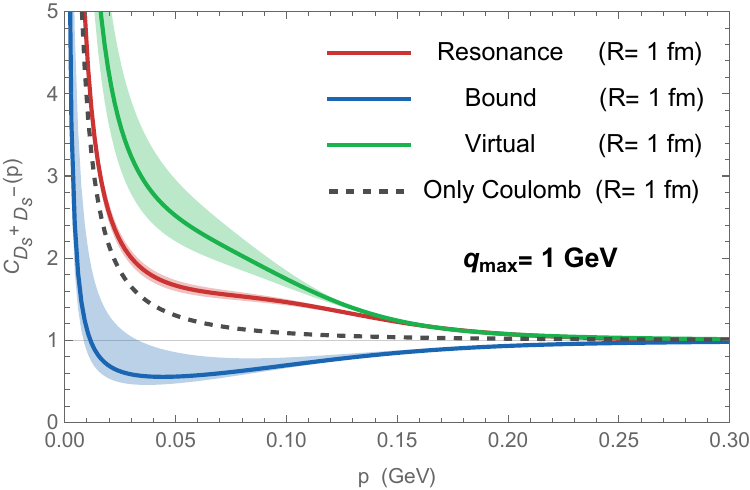}
    \label{fig_CF_total_1fm_1GeV}
    }
    \subfigure[]{
    \includegraphics[width=0.45\textwidth]{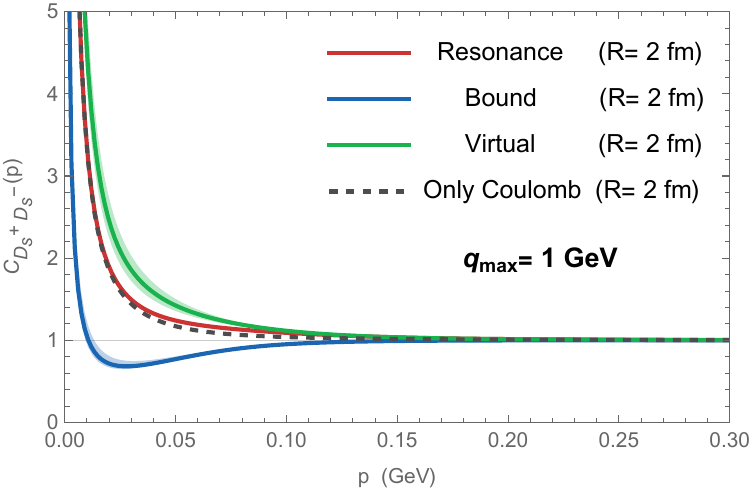}
    \label{fig_CF_total_2fm_1GeV}
    }
    \caption{\ref{fig_CF_total_1fm_0.5GeV} - \ref{fig_CF_total_2fm_1GeV}: the total $  D_s^+ D_s^- $ CF as a function of the CM relative momentum $k$, taking different values of the size parameter $R$ in three scenarios. The shaded areas represent the uncertainties in the LECs displayed in Table~\ref{table_scatprob} for the different scenarios.}
    \label{fig_CF_total}
\end{figure}

The complete $D_s^+ D_s^-$ CF, incorporating both the strong and Coulomb interactions, is presented in Fig.~\ref{fig_CF_total} as a function of the relative momentum $k$ for different source sizes $R$ and for each of the three scenarios.
A significant enhancement is observed at the $D_s^+ D_s^-$ threshold in all cases, due to the attractive
Coulomb force. Notwithstanding, the low-momentum region retains a distinct and characteristic behavior across interpretations.
For small source sizes, the differences are most pronounced: the virtual-state scenario produces a strong enhancement, the bound-state scenario leads to a clear suppression, and the resonant configuration yields a moderate augmentation of the CF, relative to the pure Coulomb CF.
At a relative momentum of $k \approx 100$ MeV, the CF exhibits significant deviations from unity: an enhancement of $\sim 75\%$ for the virtual state, a smaller enhancement of $\sim 15\%$ for the resonance, and a suppression of $\sim 40\%$ for the bound state. These pronounced deviations demonstrate the combined effects of Coulomb and strong interactions.
Our results indicate that these distinct patterns persist for smaller values of $R$, corresponding to smaller collision systems such as $pp$ and $pA$ with light nuclei. Furthermore, they exhibit less sensitivity to the choice of $q_{\text{max}}$, as evidenced in Figs.~\ref{fig_CF_strong} and \ref{fig_CF_total} for values of 0.5 GeV and 1.0 GeV. This behavior underscores the importance of measuring the $D_s^+ D_s^-$ CF in small collision systems, where the contrast between the interpretations is most pronounced.

Hence, the use of femtoscopy to distinguish between a bound, virtual, and resonant state in the $D_s^+ D_s^-$ system constitutes the central result of this work, providing a valuable tool to clarify the nature of the $X(3960)$ state.

\section{Conclusions}

In this work, we have given another example of how the combined analysis of lineshapes and femtoscopic correlations can help discriminate the underlying structure of exotic hadronic states. In particular, we have investigated the controversy concerning the nature of the $X(3960)$ state, interpreting it as a near-threshold resonance, a virtual state, or a bound $D_s^+ D_s^-$ molecule. 
Using the Bethe-Salpeter formalism to parameterize the strong interaction, our analysis finds that all three scenarios can reproduce the $D_s^+D_s^-$ invariant mass spectrum of $B^+ \to D_s^+D_s^-K^+$ decays within current experimental uncertainties. 
After that, by including Coulomb effects in the formalism, we have shown that the $D_s^+ D_s^-$ correlation function exhibits distinctive behavior in the low-momentum region, allowing it to distinguish between the three scenarios. Specifically, we predict: a strong enhancement for the virtual-state interpretation, a moderate enhancement for the resonant case, and a clear suppression for the bound-state configuration. These distinct signatures are most pronounced for small source sizes, corresponding possibly to $pp$ and $pA$ collision systems.

Some notes are warranted to discuss the consistency of the findings reported above, particularly regarding coupled-channel effects. As mentioned in the Introduction, a coupled-channel analysis of $D\bar{D}$ and $D_s\bar{D}_s$ systems performed in Ref.~\cite{Bayar:2022dqa} found two poles: one below the $D\bar{D}$ threshold and another at $3932.72$ MeV with a width of $12.32$ MeV, identified as the $X(3930)$. 
This state couples more strongly to $D_s\bar{D}_s$ than to $D\bar{D}$ and lies closer to the $D_s\bar{D}_s$ threshold than the pole position of our bound state scenario (Table~\ref{table_scatprob}).
We have checked that the $D_s\bar{D}_s$ CF obtained using the strong scattering amplitude of Ref.~\cite{Bayar:2022dqa} does not show qualitative differences with respect to that presented in the previous section for the bound state configuration. 

To conclude, given the recent advances in heavy-flavor femtoscopy, measurements of $D_s^+ D_s^-$ correlations in high-multiplicity $pp$, $pA$, and $AA$ collisions might be feasible in the near future. They would provide the experimental input needed to shed light on the nature of the $X(3960)$, thereby establishing femtoscopy as a relevant tool in studies of exotic hadrons.

\begin{acknowledgments}
This work is partly supported by the National Key R\&D Program of China under Grant No. 2023YFA1606703 and the National Natural Science Foundation of China under Grant No. 12435007.
L.M.A. would like to thank the hospitality of Beihang University. 
L.M.A. acknowledges the financial support by the Brazilian CNPq (Grants No. 400215/2022-5, 308299/2023-0, 402942/2024-8) and CNPq/FAPERJ under the Project INCT-F\'{\i}sica Nuclear e Aplica\c c\~oes (Contract No. 464898/2014-5 and 408419/2024-5). Zhi-Wei Liu acknowledges support from the National Natural Science Foundation of China (12405133, 12347180), China Postdoctoral Science Foundation (2023M740189), and the Postdoctoral Fellowship Program of CPSF (GZC20233381).

\end{acknowledgments}

\appendix

\bibliographystyle{apsrev4-1}
\bibliography{CF-DsDsBar}

\end{document}